\documentclass[prl,twocolumn,superscriptaddress,preprintnumbers,a4paper,amsmath,amssymb]{revtex4}

\usepackage{graphicx}
\usepackage{dcolumn}
\usepackage{bm}

\begin{document}

\newcommand{\zz}{3\textit{d}$_{3z^{2}-r^{2}}$}
\newcommand{\xx}{3\textit{d}$_{3x^{2}-r^{2}}$}
\newcommand{\yy}{3\textit{d}$_{3y^{2}-r^{2}}$}
\newcommand{\xy}{3\textit{d}$_{x^{2}-y^{2}}$}

\newcommand{\app}{\mbox{$\sim$}} 
\newcommand{\sinth}{\mbox{$\sin\theta/\lambda$}} 
\newcommand{\inA}{\mbox{\AA$^{-1}$}} 
\newcommand{\mub}{\mbox{$\mu_{B}$}} 
\newcommand{\mns}{$-$} 
\newcommand{\ddd}{3\textit{d}} 
\newcommand{\pp}{2\textit{p}} 
\newlength{\minusspace} 
\settowidth{\minusspace}{$-$} 
\newcommand{\msp}{\hspace*{\minusspace}} 
\newlength{\zerospace} 
\settowidth{\zerospace}{$0$} 
\newcommand{\zsp}{\hspace*{\zerospace}} 
\newcommand{\mnt}{$Mn^{3+}$} 
\newcommand{\mnf}{$Mn^{4+}$} 
\newcommand{\eg}{$\textit{e}_{g}$} 
\newcommand{\tg}{$\textit{t}_{2g}$} 
\newcommand{\q}{$\delta$}
\newcommand{\ql}{$\delta_{l}$}
\newcommand{\qlmn}{$\delta_{l}^{Mn}$}
\newcommand{\qm}{$\delta_{m}$}
\newcommand{\qmn}{$\delta_{m}^{Mn}$} 
\newcommand{\qtb}{$\delta_{m}^{Tb}$} 
\newcommand{\qc}{$\delta_{m}=\frac{1}{4}$} 
\newcommand{\qcs}{$\delta_{l}=\frac{1}{2}$} 
\newcommand{\Tnmn}{T$_{N}^{Mn}$} 
\newcommand{\Tntb}{T$_{N}^{Tb}$} 
\newcommand{\Tn}{T$_{N}$} 
\newcommand{\TN}{T$_{N}$} 
\newcommand{\Tl}{T$_{lock}$} 
\newcommand{\etal}{\textit{et al.}} 
\newcommand{\degc}{$^{\circ}$C} 
\newcommand{\degg}{$^{\circ}$} 
\newcommand{\at}{$A^{\prime}$}
\newcommand{\ct}{$C^{\prime}$}
\newcommand{\ft}{$F^{\prime}$}
\newcommand{\gt}{$G^{\prime}$}
\newcommand{\ha}{$H\|a$} 
\newcommand{\hb}{$H\|b$} 
\newcommand{\hc}{$H\|c$} 
\newcommand{\pa}{$P\|a$} 
\newcommand{\pb}{$P\|b$} 
\newcommand{\pc}{$P\|c$} 
\newcommand{\mc}{$\mu C/m^{2}$} 
\newcommand{\Tc}{T$_{\mathrm C}$} 
\newcommand{\Tic }{$T_{C}$} 
\newcommand{\Hic }{$H_{IC-C}$} 
\newcommand{\Haic }{$H\|a_{IC-C}$} 
\newcommand{\Hbic }{$H\|b_{IC-C}$} 
\newcommand{\Hcb }{$H^{b}_{C}$} 
\newcommand{\Hca }{$H^{a}_{C}$} 
\newcommand{\Hc }{$H_{C}$} 
\newcommand{\arrows}{$\uparrow \uparrow \downarrow \downarrow$}

\preprint{}

\title{Field induced linear magneto-elastic coupling in multiferroic TbMnO$_{3}$}

\author{N. Aliouane} 
\affiliation{Hahn-Meitner-Institut, Glienicker Str. 100, Berlin D-14109, 
Germany} 
\author{D. N. Argyriou} 
\affiliation{Hahn-Meitner-Institut, Glienicker Str. 100, Berlin D-14109, 
Germany} 
\author{J. Strempfer}
\author{I. Zegkinoglou}
\affiliation{Max-Planck-Institut f\"ur Festk\"orperforschung,Heisenbergstra\ss e 1, 70569 Stuttgart, Germany}
 \author{S. Landsgesell}
\affiliation{Hahn-Meitner-Institut, Glienicker Str. 100, Berlin D-14109, Germany}
\author{M. v. Zimmermann}
\affiliation{Hamburger Synchrotronstrahlungslabor HASYLAB at Deutsches Elektronen-Synchrotron DESY, Notkestr. 85, 22603 Hamburg, Germany}
\date{\today}
\preprint{}
\begin{abstract}
We have used in-field neutron and X-ray single crystal diffraction to measure the incommensurability $\delta$ of the crystal and magnetic structure of multiferroic TbMnO$_{3}$. We show that the flop in the electric polarization at the critical field $H_{C}$,  for field $H$ along the $a-$ and $b-$axis coincides with a 1st order transition to a commensurate phase with propagation vector $\kappa=(0,\frac{1}{4},0)$.  In-field X-ray diffraction measurements show that the quadratic magneto-elastic coupling breaks down with applied field as shown by the observation of the 1st harmonic lattice reflections above and below $H_{C}$. This indicates that magnetic field induces a linear magneto-elastic coupling. We argue that the  commensurate phase can be described by an ordering of Mn-O-Mn bond angles. 
\end{abstract}


\maketitle

Control of the spontaneous ferroelectric polarization ($P_{s}$) with an external magnetic field ($H$) in a material opens the opportunity for new types of magneto-electric devices. The realization of such devices is based on multiferroic materials in which magnetism and ferroelectricity are strongly coupled. While available multiferroics  are limited, it has been shown that  frustrated spin materials may offer a new class of enhanced multiferroics \cite{kimura,kimura2,goto}. In one of these new materials, TbMnO$_{3}$, we find that multiferroic behavior arises as a consequence of the release of frustration with $H$. Here ferroelectricity arises  below the N\'eel temperature (\Tn), from a coupling to the lattice of an incommensurate (IC) modulation of the magnetic structure (Fig. \ref{strt}a) that is caused from frustration in the ordering Mn $d$ orbitals \cite{kimura,kimura2}. In this letter we show that magnetic field releases this frustration,  inducing a linear magneto-elastic coupling, so that ferroelectricity is no longer a secondary effect. The linear magneto-elastic coupling   drives a magneto-structural transition from an IC phase which has $P_{s}$ along the $c-$axis to a commensurate (C) phase with $P_{s}$ along the $a$-axis.  

In TbMnO$_{3}$ (orthorhombic, $a$=5.302\AA, $b$=5.857\AA, $c$=7.402\AA) when a magnetic field is applied along the $b$-axis (\hb) at 2K, parallel to the direction of the  IC magnetic modulation (see Fig. \ref{strt}a),  the electric polarization of the lattice flops from along the $c-$axis (\pc) to along the $a-$axis (\pa) at the critical field  \Hcb\app4.5T (Fig.~\ref{polq}(d)). When field is applied along the $a$-axis (\ha), perpendicular to the  magnetic modulation, a similar  flop is found but at a higher critical field, \Hca\app 9T (Fig.~\ref{polq}(a)). Recently there have been a number of examples of novel magneto-elastic coupling in complex multiferroic oxides such as Tb$_{2}$Mn$_{2}$O$_{5}$ which exhibits a reversable polarization switch with applied field \cite{hur}, and hexagonal HoMnO$_{3}$ where one  magnetic phase is selected over another by applying an electric field \cite{lotter}. However, TbMnO$_{3}$ is unique as it is the only known example of a material that exhibits a field induced flop of the polarization.

In TbMnO$_{3}$ the staggered ordering of Mn$^{3+}$ \xx /\yy\space orbitals as found in LaMnO$_{3}$ is frustrated partly due to the small ionic size of Tb$^{3+}$ \cite{kimura2}. This leads to an IC spin ordering which drives a ferroelectric lattice modulation \cite{kimura2, kenz}.  The wavevector for the modulation of the Mn magnetic moments  is $\kappa_{m}^{Mn}=(0,k\pm$\qmn$,0)$, with incommensurability \qmn$\sim$0.29 at \TN\space(1st harmonic) \cite{Quezel,kajimoto}.  Accompanying the magnetic ordering there is  a lattice modulation with \ql=2\qm\space (2nd harmonic) \cite{goto,kimura,kimura2} consistent with a \emph{quadratic magneto-elastic coupling} between the lattice and the spin density wave \cite{PW,walker}.  Below \Tntb=7K, sinusoidal Tb spin order is also found ($\kappa_{m}^{Tb}=(0,k\pm$\qtb$,0)$, \qtb=0.42) \cite{kajimoto} and is expected to produce a similar coupling to the lattice with $\delta_{m}^{Tb}=2\delta_{l}^{Tb}$. 

To uncover the nature of the coupling between $P$ and $H$ we have used in-field neutron and synchrotron X-ray single crystal diffraction to measure the field response of $\delta$ with field applied parallel to the $a-$ and $b-$axis (\ha\space and \hb), up to 14T. Single crystals  were obtained by re-crystalizing  a ceramic rod of TbMnO$_{3}$ under Ar atmosphere using an optical floating zone furnace. Small pieces of our crystal were characterized using magnetization and specific heat with good agreement with published measurements \cite{kimura}. Synchrotron X-ray diffraction measurements were carried out on beamline X21 ($E_{i}=$9.5 keV) at the NSLS, Brookhaven National Laboratory, using a vertical 13T superconducting magnet with the sample alighed with \ha, and at the BW5 beamline at HASYLAB ($E_{i}=$100.5 keV), using a horizontal 10T superconducting magnet and the sample aligned with \hb. Neutron diffraction experiments were made on  a single crystal of TbMnO$_{3}$, (19mm in length and 8mm in diameter)  at the BENSC facility of the Hahn-Meitner Institut, Berlin. For measurements with \ha\space we used  the FLEX cold triple-axis spectrometer with collimation of 40$'$-40$'$-40$'$, $k_{i}$=1.5\AA$^{-1}$, and a cooled Be filter on the scattered beam, while a magnetic field was applied using the vertical 14.5T superconducting magnet VM1.  For measurements with \hb\space we used  the 2-axis diffractometer E4 ($\lambda=2.2$\AA) with 60$'$-40$'$ collimation while field was applied using the 4T horizontal field magnet HM1. In all cases diffraction measurements were made in the $bc$-plane. Our neutron diffraction measurements are sensitive to magnetic ordering whereas X-ray diffraction experiments probe the lattice modulation. 

We applied field parallel to the $a-$axis and we measured characteristic reflections using neutron diffraction by cooling our crystal from 50 to 2K in fields of 0,2,4,6,10 and 14T.  The variation of $\delta$ for Mn and Tb spin ordering as a function of field at 2K, compiled from these measurements, is shown in Fig. \ref{polq}(b), while the variation of the intensity of the same reflections is shown in Fig. \ref{polq}(c). 

In Fig. \ref{flex}(b) we show neutron diffraction scans through the 1st harmonic reflection $(0,$\qmn$,1)$ as a function of \ha. The intensity of this reflection is enhanced with magnetic field from 2 to 4T while the magnitude of \qmn\space remains unchanged for fields  up to 8T (Fig.\ref{polq}(b,c)). However, for $H>$\Hca$\sim$9T we find that $\delta$ of the 1st harmonic reflection has changed to a C value of \q=$\frac{1}{4}$ and is accompanied by a 2nd harmonic reflection with 2\q=$\frac{1}{2}$ (Fig. \ref{polq}(b)). This transition at \Hca\space coincides  with the polarization flop from \pc\space to \pa\space as shown in Fig. \ref{polq}(a). The implication of this observation is that the polarization flop arises because the spontaneous polarization of the C-phase is along the $a-$axis, different to that of the IC-phase. 

In Fig. \ref{flex}(a) we show X-ray diffraction scans as a function of \ha\space at T=10K for the 1st harmonic  (0,3-$\delta^{Mn}$,8) reflection.  If the magneto-elastic coupling is quadratic, then only magnetic scattering is allowed for this reflection, and indeed we do not observe a signal at this position for \ha$<$2T.  However, for \ha$\geq$2T  a peak becomes visible and above \Hca\space it locks into the C value of \q=$\frac{1}{4}$. The observation of a lattice reflection at the wavevector of the 1st harmonic indicates that for \ha\space a linear magneto-elastic coupling is induced above 2T which is maintained above \Hc\space into the C-phase.  The variation of the intensity of this reflection, shown in Fig. \ref{polq}(c) is complex, but its changes are associated with the metamagnetic transition of Tb-spins (see below)  and the increase in the intensity of the 1st harmonic reflection $(0,$\qm$,1)$ between 2 and 4T (see Fig. \ref{polq}(c)).

The behavior of the magnetic reflections due to Tb ordering is significantly different from that for Mn. The intensity of the 1st harmonic reflection $(0,$\qtb$,1)$ decreases significantly between 0 and 2T (Fig. \ref{polq}(c)), indicative of a metamagnetic transition reported previously \cite{kimura}. This transition in the Tb magnetic sub-lattice is coincident with an anomaly in \pc\, as well as with a sharp increase in the magnetization reported elsewhere \cite{kimura,kimura3}.  However, above \Hca\space we find no evidence of sinusoidal Tb spin ordering, suggesting that Tb spins have also ordered commensurately. Rather we find that with \ha\space Tb spins are polarized ferromagneticaly as indicated by an increase in the intensity of Bragg reflections. 

The polarization flop when the field is applied along the $b-$axis in the direction of the magnetic modulation, occurs at a lower critical field of \Hcb$\sim$4T (Fig. \ref{polq}(d)). Here  
we applied a magnetic field parallel to the $b-$axis after the sample was cooled to 2K in zero field and characteristic superlattice reflections, such as the $(0,2\delta^{Mn},3)$, were measured using synchrotron X-ray diffraction (see Fig. 4). When a magnetic field is applied, the incommensurability of the 2nd harmonic reflection  remains invariant (Fig. \ref{polq}(e)) but its intensity rapidly decreases with field from 0 to 2T and remains low up to $\sim$4T (
Fig. \ref{polq}(f)).  Increasing field clearly destabilizes the lattice modulation and leads to a discontinuous transition to the C-phase with 2\q=$\frac{1}{2}$  at \Hcb=4.5T (Fig. \ref{hasy}(a) and \ref{polq}(e,f)).  We find C and IC reflections coexisting over the field range of 4 to 5.5T, while for  $H>$5.5T only the  2\q=$\frac{1}{2}$ superlattice reflection is visible (Fig. \ref{hasy}(a)). Coincident to this magneto-structural  transition is the polarization flop from \pc\space to \pa\space at about 4.5T as shown in Fig.~\ref{polq}(d). 
 
As with the previous field configuration, here also our X-ray measurements revealed superlattice reflections with \q=$\frac{1}{4}$ above \Hcb\space (Fig. \ref{hasy}(b) and \ref{polq}(e,f)). The linear increase in the intensity of this reflection with \hb\space above \Hcb\space is particularly striking, as it suggests a linear magneto-electric coupling in the C-phase. The variation of intensity with field of the  2nd harmonic reflection is different in that it increases rapidly above \Hcb\space and shows a constant variation with $H$ above $\sim$6T (Fig. \ref{polq}(f)). 

For this field configuration we find using neutron diffraction that at 2K Tb spins undergo an additional discontinuous transition to a phase with \qtb=$\frac{1}{3}$ at \hb=1.25T, as shown in Fig.~\ref{polq}(e). From symmetry analysis Tb-spins for this phase are both sign and amplitude modulated. This transition is reflected in an anomaly in \pc\space at the same field value (Fig.~\ref{polq}(d)), and a significant increase in the magnetization at this temperature is also reported\cite{kimura}. Above \Hcb\space we find no evidence of incommensurate ordering of Tb-spins, but the significant increase of the magnetization at this field indicates that Tb spins order commensurately there \cite{kimura}.

In zero field TbMnO$_{3}$ is an improper ferroelectric where the lattice polarization  arises from a quadratic magneto-elastic coupling to a spin density wave as indicated by the observation of the 2nd harmonic structural reflections \cite{PW,walker}. Here we demonstrate that this picture of improper ferroelectricity in TbMnO$_{3}$ breaks down with $H$ even below  $H_{C}$ for the case of \ha\space and most likely for \hb. The identical periodicity of the magnetic and crystal structure indicates that $H$ switches on a linear magneto-elastic coupling which is maintained into the C-phase. This would suggest that the IC-C phase transition we report on and consequently the polarization flop at $H_{C}$ are driven by field induced linear magneto-elastic coupling. Further our measurements show that although the application of field ultimately results in a C-phase, the effect of $H$ on Tb-spins is dependent on the field direction. For \hb\space the transition to the C-phase is coupled to a metamagnetic transition of Tb-spins \cite{kimura}, while for \ha\space the Tb-spins order commensurately at field below \Hcb. How exactly Tb-spins couple to the transition of the high field C-phase is yet to be resolved. 

The \emph{magneto-electric} coupling is expected to change in a similar way as the magneto-elastic coupling.  Indeed the  behavior of TbMnO$_{3}$ with $H$ is reminiscent to that of BiFeO$_{3}$ which exhibits a switch  to linear magneto-electric coupling at $\sim$20T\cite{popov}. The coupling between $P$ and $H$ is expressed by $P_{i}=P_{si}+\alpha_{ij}H_{j}+\frac{1}{2}\beta_{ijk}H_{j}H_{k}$, where $P_{s}$ is the spontaneous polarization,  $\alpha_{ij}$ is the tensor of the magneto-electric susceptibility, $\beta_{ijk}$ is a tensor describing the coupling of an incommensurate magnetic structure to the lattice 
\cite{popov}. In an analogy to BiFeO$_{3}$, we suggest that as field destroys the incommensurate spin ordering ($\beta_{ijk}\rightarrow$0), a linear magneto-electric effect would be expected to arise \cite{popov,zv}. Indeed a linear behavior in $P$ vs. $H$ is evident for \hb\space(Fig. 2(d)).  

The commensurate propagation vector of $\kappa=(0,\frac{1}{4},0)$ uniquely gives a spin structure that is only sign modulated as depicted in Fig. \ref{strt}(b). It is of interest to consider the effect of this magnetic ordering on the 
frustrated orbital ordering. The crystal structure of  TbMnO$_{3}$ (H=0T, $T>$\TN)  shows an ordering of long and short bonds as found in LaMnO$_{3}$, but with a low angle $\phi$ of 144\degg\space between MnO$_{6}$ octahedra (Mn-O-Mn), compared to 165\degg\space found in LaMnO$_{3}$\cite{blasco}. A simple application of  Goodenough's rules would suggest a layered antiferromagnetic (AF) ordering, inconsistent with the present data\cite{Goodenough}.   We propose here that the structural distortion that is coupled to the \arrows\space spin structure arises from the ordering of $\phi$. Small values of $\phi$ would result in AF interactions between adjacent Mn-spins as the influence of super-exchange via \tg-orbitals would be more significant \cite{Goodenough,kimura2}.  On the other hand a relative larger value of $\phi$ will be less influenced from \tg\space interactions and would propagate a ferromagnetic coupling between Mn-spins as it does in LaMnO$_{3}$. Indeed the layered AF structure is stable to values of $\phi$ as low as 148\degg,  found in SmMnO$_{3}$ \cite{kimura2}.  Using the magnetic ordering of the high field C-phase as a constraint and displacing O-atoms in the para-electric structure so that $\phi$ should increase (decrease) for ferromagnetic (antiferromagnetic) coupling between adjacent Mn-spins, one arrives at a structural model show in Fig. \ref{strt}b. The result is a structure were ferromagnetically coupled MnO$_{6}$ octahedra undergo simple rotations (ferroelectrically inactive), while octahedra that are frustrated by antiferromagnetic coupling in the $ab$-plane show a scissor-like distortion (ferroelectrically active).  The combined  rotations and scissor-like distortions in the structure leads to anti-ferroelectric displacements for the O-atoms along the $b-$axis but an overall ferroelectric displacement  along the $a-$axis in agreement with polarization measurements. While this simple model predicts the correct polarization for the C-phase on the basis of the magnetic structure, the periodicity of the structural distortion suggests displacements of the Mn ions that can not be accurately predicted in the absence of high field crystallographic data.  
 
In this letter we have demonstrated that the polarization flop in TbMnO$_{3}$ arises from a transition to a C-phase with propagation vector $\kappa=(0,\frac{1}{4},0)$.  The quadratic magneto-elastic coupling breaks down with field and the high field phase exhibits linear magneto-elastic and magneto-electric behavior. We argue that field has the effect of releasing the apparent frustration in the orbital ordering in TbMnO$_{3}$ and leads to a C-phase  by an ordering of Mn-O-Mn bond angles. 
 
\begin{acknowledgments}
The authors benefited from discussions with  D. Khomskii, T. Kimura and L.C. Chapon.  We thank Klaus Habicht, Peter Smeibidl, Sebastian Gerischer, Klaus Kiefer, and Michael Meissner from BENSC, C.J. Milne from HMI and W.A. Caliebe from NSLS for assistance during the experimental work.Work at Brookhaven was supported by the
U.S. Department of Energy, Division of Materials Science, under
Contract No. DE-AC02-98CH10886.
\end{acknowledgments}


\begin{thebibliography}{15}
\expandafter\ifx\csname natexlab\endcsname\relax\def\natexlab#1{#1}\fi
\expandafter\ifx\csname bibnamefont\endcsname\relax
  \def\bibnamefont#1{#1}\fi
\expandafter\ifx\csname bibfnamefont\endcsname\relax
  \def\bibfnamefont#1{#1}\fi
\expandafter\ifx\csname citenamefont\endcsname\relax
  \def\citenamefont#1{#1}\fi
\expandafter\ifx\csname url\endcsname\relax
  \def\url#1{\texttt{#1}}\fi
\expandafter\ifx\csname urlprefix\endcsname\relax\def\urlprefix{URL }\fi
\providecommand{\bibinfo}[2]{#2}
\providecommand{\eprint}[2][]{\url{#2}}

\bibitem[{\citenamefont{Kimura et~al.}(2003{\natexlab{a}})\citenamefont{Kimura,
  Goto, Shintani, Ishizaka, Arima, and Tokura}}]{kimura}
\bibinfo{author}{\bibfnamefont{T.}~\bibnamefont{Kimura}},
 \bibnamefont{et~al.},
   \bibinfo{journal}{Nature} \textbf{\bibinfo{volume}{426}}, \bibinfo{pages}{55}
  (\bibinfo{year}{2003}{\natexlab{a}}),

\bibitem[{\citenamefont{Goto et~al.}(2004)\citenamefont{Goto, Kimura, Lawes,
  Ramirez, and Tokura}}]{goto}
\bibinfo{author}{\bibfnamefont{T.}~\bibnamefont{Goto}},
\bibnamefont{et~al.},
   \bibinfo{journal}{Phys. Rev. Lett.} \textbf{\bibinfo{volume}{92}},
  \bibinfo{eid}{257201} (\bibinfo{year}{2004}),
 
\bibitem[{\citenamefont{Kimura et~al.}(2003{\natexlab{b}})\citenamefont{Kimura,
  Ishihara, Shintani, Arima, Takahashi, Ishizaka, and Tokura}}]{kimura2}
\bibinfo{author}{\bibfnamefont{T.}~\bibnamefont{Kimura}},
\bibnamefont{et~al.},
  \bibinfo{journal}{Phys. Rev. B}
  \textbf{\bibinfo{volume}{68}}, \bibinfo{eid}{060403}
 (\bibinfo{year}{2003}{\natexlab{b}}),

\bibitem[{\citenamefont{Hur et~al.}(2004)\citenamefont{Hur, Park, Sharma, Ahn,
  Guha, and Cheong}}]{hur}
\bibinfo{author}{\bibfnamefont{N.}~\bibnamefont{Hur}},
\bibnamefont{et~al.},
  \bibinfo{journal}{Nature} \textbf{\bibinfo{volume}{429}},
  \bibinfo{pages}{392} (\bibinfo{year}{2004}),

\bibitem[{\citenamefont{Lottermoser et~al.}(2004)\citenamefont{Lottermoser,
  Lonkai, Amann, Hohlwein, Ihringer, and Fiebig}}]{lotter}
\bibinfo{author}{\bibfnamefont{T.}~\bibnamefont{Lottermoser}},
\bibnamefont{et~al.},
  \bibinfo{journal}{Nature} \textbf{\bibinfo{volume}{430}},
  \bibinfo{pages}{541} (\bibinfo{year}{2004}).

\bibitem[{\citenamefont{Kenzelmann et~al.}(2005)\citenamefont{Kenzelmann,
  Harris, Jonas, Broholm, Schefer, Kim, Zhang, Cheong, Vajk, and Lynn}}]{kenz}
\bibinfo{author}{\bibfnamefont{M.}~\bibnamefont{Kenzelmann}},
\bibnamefont{et~al.},
  \bibinfo{journal}{Phys. Rev. Lett.} \bibinfo{}{in press}
  (\bibinfo{year}{2005}).

\bibitem[{\citenamefont{Quezel et~al.}(1977)\citenamefont{Quezel, Tcheou,
  Rossat-Mignod, Quezel, and Roudaut}}]{Quezel}
\bibinfo{author}{\bibfnamefont{S.}~\bibnamefont{Quezel}},
\bibnamefont{et~al.},
  \bibinfo{journal}{Physica B+C} \textbf{\bibinfo{volume}{86-88}},
  \bibinfo{pages}{916} (\bibinfo{year}{1977}),

\bibitem[{\citenamefont{Kajimoto et~al.}(2004)\citenamefont{Kajimoto,
  Yoshizawa, Shintani, Kimura, and Tokura}}]{kajimoto}
\bibinfo{author}{\bibfnamefont{R.}~\bibnamefont{Kajimoto}},
\bibnamefont{et~al.},
  \bibinfo{journal}{Phys. Rev. B}
  \textbf{\bibinfo{volume}{70}}, \bibinfo{eid}{012401}
   (\bibinfo{year}{2004}),

\bibitem[{\citenamefont{Plumer and Walker}(1982)}]{PW}
\bibinfo{author}{\bibfnamefont{M.~L.} \bibnamefont{Plumer}} \bibnamefont{and}
  \bibinfo{author}{\bibfnamefont{M.~B.} \bibnamefont{Walker}},
  \bibinfo{journal}{Journal of Physics C: Solid State Physics}
  \textbf{\bibinfo{volume}{15}}, \bibinfo{pages}{7181} (\bibinfo{year}{1982}).

\bibitem[{\citenamefont{Walker}(1980)}]{walker}
\bibinfo{author}{\bibfnamefont{M.~B.} \bibnamefont{Walker}},
  \bibinfo{journal}{Phys. Rev. B}
  \textbf{\bibinfo{volume}{22}}, \bibinfo{pages}{1338} (\bibinfo{year}{1980}),


\bibitem[{\citenamefont{Kimura et~al.}((2005))\citenamefont{Kimura, Lawes,
  Goto, Tokura, and Ramirez}}]{kimura3}
\bibinfo{author}{\bibfnamefont{T.}~\bibnamefont{Kimura}},
\bibnamefont{et~al.},
  \bibinfo{journal}{Phys. Rev. B}  \bibinfo{pages}{in press}
  (\bibinfo{year}{(2005)}).
  
\bibitem[{\citenamefont{Popov et~al.}(1993)\citenamefont{Popov, Zvezdin,
  Vorob'ev, Kadomtseva, Murashev, and Rakov}}]{popov}
\bibinfo{author}{\bibfnamefont{Y.}~\bibnamefont{Popov}},
\bibnamefont{et~al.},
  \bibinfo{journal}{JETP Lett.} \textbf{\bibinfo{volume}{57}},
  \bibinfo{pages}{69} (\bibinfo{year}{1993}).

\bibitem[{\citenamefont{Zvezdin and Pyatakov}(2004)}]{zv}
\bibinfo{author}{\bibfnamefont{A.~K.} \bibnamefont{Zvezdin}} \bibnamefont{and}
  \bibinfo{author}{\bibfnamefont{A.~P.} \bibnamefont{Pyatakov}},
  \bibinfo{journal}{Phys. Usp.} \textbf{\bibinfo{volume}{47}},
  \bibinfo{pages}{416} (\bibinfo{year}{2004}).


\bibitem[{\citenamefont{Blasco et~al.}(2000)\citenamefont{Blasco, Ritter,
  Garcia, de~Teresa, Perez-Cacho, and Ibarra}}]{blasco}
\bibinfo{author}{\bibfnamefont{J.}~\bibnamefont{Blasco}},
\bibnamefont{et~al.},
 \bibinfo{journal}{Phys. Rev. B} \textbf{\bibinfo{volume}{62}}, \bibinfo{pages}{5609}
  (\bibinfo{year}{2000}).
  
\bibitem[{\citenamefont{Goodenough}(1955)}]{Goodenough}
\bibinfo{author}{\bibfnamefont{J.}~\bibnamefont{Goodenough}},
  \bibinfo{journal}{Phys. Rev.} \textbf{\bibinfo{volume}{100}},
  \bibinfo{pages}{564} (\bibinfo{year}{1955}).

\end{thebibliography}


\begin{figure}[bt!]
\begin{center}
\caption{Crystal and magnetic structure of TbMnO$_{3}$ in the $ab-$plane. Long (darker) and short (lighter) Mn-O bond ordering  arises from staggered \xx/\yy\space orbital ordering found above \TN. (a) A Model of the IC magnetic structure of TbMnO$_{3}$ below \TN\space\cite{Quezel}. Arrows through Mn atoms indicate the direction of Mn-spins. (b) A model of the Mn-spin structure for TbMnO$_{3}$ for H$>$\Hc. Here arrows from O-atoms indicate the direction of displacement from the average structure as to increase (decrease) the angle $\phi$ on the basis of ferromagnetic (antiferromagnetic) coupling between adjacent Mn-ions. Large arrows on ferroelectrically active octahedra indicate the direction of the predicted polarization from these displacements.}
\label{strt}
\end{center}
\end{figure}

\newpage

\begin{figure}[t!]
\begin{center}
\caption{(a,d) Electric polarization of TbMnO$_{3}$ parallel to the $a$- (\pa) and $c$-axis (\pc) with \ha\space and \hb, respectively. These data were compiled from field cooled measurements reported by Kimura \etal\cite{kimura3}. (b) Variation of  $\delta$ as a function of \ha, for the 1st harmonic reflections $(0,$\qmn$,1)$, $(0,$\qtb,$1)$ measured using neutron diffraction (filled blue and red circles respectively) and $(0,3-$\q$,8)$ measured with X-ray diffraction (filled green squares). For $H>$\Hca\space the C reflections  $(0,\frac{1}{4},1)$ and $(0,\frac{1}{2},1)$ (open blue circles) were measured using neutron diffraction while the  $(0,2\frac{3}{4},8)$ reflection with X-rays (open green squares). (c) Variation of the integrated intensity as a function \ha, for the same reflections as in panel (b). For the reflection due to Tb ordering the logarithmic scale on the right axis is used. (e) Variation of $\delta$ as a function  \hb, for the 1st harmonic reflections $(0,$\qtb,$1)$ and C $(0,\frac{1}{3},1)$
due to Tb spin ordering (filled and unfilled red circles respectively). On the same panel we show the results of X-ray diffraction measurements of the 2nd harmonic Mn reflection $(0,2\delta^{Mn},3)$ (filled blue circles) and the 1st and 2nd harmonic reflections $(0,\frac{1}{2},3)$ and $(0,\frac{1}{4},3)$ (open blue circles). (f). Variation of the integrated intensity as a function of \hb, for the reflections shown in panel (e). For the reflection due to Tb ordering the logarithmic scale on the right axis is used. Integrated intensity in panels (c) and (d) is shown in arbitrary units.}
\label{polq}
\end{center}
\end{figure}

\begin{figure}[t!]
\caption{(a) X-ray diffraction scans of the 1st harmonic  (0,3-$K$,8) reflection as a function of field with \ha\space at T=10K, measured on X21.  (b) Neutron diffraction scans through the 1st harmonic (0,\qmn,1) reflection as a function of \ha. }
\label{flex}
\end{figure}

\newpage

\begin{figure}[h!]
\begin{center}
\caption{(a) Scans of the 2nd harmonic IC lattice reflection at $(0,2\delta^{Mn},3)$ with increasing field at 2.6K measured on BW5. The C reflection at (0,$\frac{1}{2}$,3) is shown for $H>$\Hcb. (b) Similarly, scans of the 1st harmonic (0,$\frac{1}{4}$,4) reflection of the C-phase with increasing field at 2.6K for $H>$\Hca\space are shown. 
}
\label{hasy}
\end{center}
\end{figure}

\end{document}